\documentclass[
12pt,
preprint,preprintnumbers,nofootinbib,
groupedaddress,superscriptaddress,amsmath,amssymb]{revtex4}
\usepackage[dvipdfmx]{graphicx}
\usepackage{dcolumn}
\usepackage{bm}
\usepackage{amssymb}
\usepackage{amsmath}
\usepackage{epsfig}    
\usepackage[dvipdfmx]{color}
\usepackage{slashed}
\usepackage{hhline}
\usepackage{hyperref}
\usepackage{soul,xcolor}

\def\be{\begin{equation}}
\def\ee{\end{equation}}
\newcommand{\bea}{\begin{eqnarray}}
\newcommand{\eea}{\end{eqnarray}}
\newcommand{\nn}{\nonumber}

\numberwithin{equation}{section}

\begin{document}

\setstcolor{blue}

\title{{A Two Loop Radiative Neutrino Model }
}
%
\author{Seungwon Baek}
\email{sbaek@korea.ac.kr}
\affiliation{Department of Physics, Korea University, Seoul 02841, Republic of Korea}
\author{Hiroshi Okada}
\email{hiroshi.okada@apctp.org}
\affiliation{Asia Pacific Center for Theoretical Physics, Pohang, Gyeongbuk 790-784, Republic of Korea}

\author{Yuta Orikasa}
\email{Yuta.Orikasa@utef.cvut.cz}
\affiliation{Institute of Experimental and Applied Physics, Czech Technical University, Prague 12800, Czech Republic}

\date{\today}

\begin{abstract}
We explore the possibility to explain a bosonic dark matter candidate with a gauge singlet inside the loop to generate the neutrino mass matrix at two-loop level. 
The mass matrix is suppressed by a small mixing that comes from the bound on {direct detection experiments of the dark matter,
and equivalent of the three-loop neutrino model due to the small mixing between neutral inert bosons.
Here, our setup is} the Zee-Babu type scenario with $Z_3$ discrete symmetry, in which we consider the neutrino oscillation data, lepton flavor violations, muon $g-2$,  $\mu-e$ conversion rate, lepton flavor-changing and conserving  $Z$ boson decay and bosonic dark matter candidate. 
\end{abstract}
\maketitle
\newpage

\section{Introduction}
\label{sec:intro}
Radiatively induced neutrino masses is one of the promising scenarios which make strong correlations between neutrinos and  any fields that
are introduced inside loops. 
If a dark matter (DM) candidate is introduced in the model, its testability
is enhanced due to the fact that parameter space is strongly constrained by neutrino data. Especially two-loop
induced models that we will focus on in this paper have been  widely
studied in various aspects \cite{2-lp-zB, Babu:2002uu, AristizabalSierra:2006gb, Nebot:2007bc, Schmidt:2014zoa,
  Herrero-Garcia:2014hfa, Long:2014fja, VanVien:2014apa, Aoki:2010ib, Lindner:2011it, Baek:2012ub, Aoki:2013gzs,
  Kajiyama:2013zla, Kajiyama:2013rla, Baek:2013fsa, Okada:2014vla, Okada:2014qsa, Okada:2015nga, Geng:2015sza,
  Kashiwase:2015pra, Aoki:2014cja, Baek:2014awa, Okada:2015nca, Sierra:2014rxa, Nomura:2016rjf, Nomura:2016run,
  Bonilla:2016diq, Kohda:2012sr, Dasgupta:2013cwa, Baek:2014sda, Nomura:2016ask, Nomura:2016pgg, Liu:2016mpf, Nomura:2016dnf}.

{In this paper, 
we study the muon anomalous magnetic moment, various lepton flavor {violations (LFVs)}, and DM phenomenology
 in the framework of Zee-Babu type of neutrino model, emphasizing $\mu-e$ conversion rate in $Titanium$ nuclei that will be tested in the near future
 experiment such as PRISM/PRIME~\cite{Barlow:2011zza}. 
{Despite a two-loop model, we will also show that the scale of neutrino mass in our model
 is equivalent to a three-loop model. It is due to the small mixing between neutral bosons
dictated by the direct detection bound of the DM candidate.}

This paper is organized as follows. 
In Sec.~II, we introduce our model,  including neutrino sector, LFVs, muon anomalous magnetic moment 
and lepton flavor-changing and conserving  $Z$  boson decay.
In Sec.~III, we present our numerical analysis and identify regions consistent with the current experiments.
We conclude and discuss in Sec.~VI.


\section{Model setup}
 \begin{widetext}
\begin{center} 
\begin{table}[t]
\begin{tabular}{|c||c|c|c||c|c|c|c|}\hline\hline  
&\multicolumn{3}{c||}{Lepton Fields} & \multicolumn{4}{c|}{Scalar Fields} \\\hline
& ~$L_L$ ~&~ $e_R^{}$ ~ &~ $N_{L/R}$ ~&~ $\Phi$ ~&~ $\eta$ ~&~ $\chi$ ~&~ $\chi^+$ ~\\\hline 
$SU(2)_L$ & $\bm{2}$  & $\bm{1}$  & $\bm{1}$ & $\bm{2}$ & $\bm{2}$ & $\bm{1}$& $\bm{1}$ \\\hline 
$U(1)_Y$ & $-\frac12$ & $-1$   & $0$ & $\frac{1}{2}$ & $\frac12$ & $0$ & $1$  \\\hline
 $Z_3$ & $1$ & $1$ & $\omega$ & $1$  & $\omega$ & $\omega$& $\omega$  \\\hline
\end{tabular}
\caption{Particle contents and charge assignments of leptons and new particles under $SU(2)_L\times U(1)_Y\times Z_3$.
Here $\omega\equiv e^{2\pi i/3}$.}
\label{tab:1}
\end{table}
\end{center}
\end{widetext}

We introduce three families of iso-spin singlet vector-like neutral fermions $N_i$ ($i=1,2,3$),
iso-spin doublet scalar $\eta$, an isospin singlet neutral scalar $\chi$ and charged scalars $\chi^\pm$ in addition to the SM fields.
We impose a discrete $Z_3$ symmetry on all the new particles $(N, \eta, \chi,\chi^\pm)$
 in order to assure the stability of DM ($\chi$ in our case). 
The particle contents and their charge assignments under $SU(2)_L \times U(1)_Y \times Z_3$ are shown in Tab.~\ref{tab:1}
\footnote{Although there are many other possible symmetries to realize our model, $Z_3$ is a minimal symmetry.}.
Thus we expect that only the SM-like Higgs $\Phi$ has a vacuum expectation value (VEV), which is denoted by
$\langle \Phi^0 \rangle =v/\sqrt2$~\cite{Aoki:2014cja}. 
Then the relevant Lagrangian and scalar potential respecting the symmetries are given by 
\begin{align}
-\mathcal{L}_{Y}
&=
(y_{\ell})_{ij} \bar L_{L_i} \Phi e_{R_j}  +(y_{\eta})_{ij} \bar L_{L_i}(i\sigma_2) \eta^* N_{R_j}  +
y_{N_{R_{ij}}} \bar N^c_{R_i} N_{R_j}\chi  +y_{N_{L_{ij}}} \bar N^c_{L_i} N_{L_j}\chi  \nn\\
&+(y_\chi)_{ij} \bar N_{L_i} e_{R_j} \chi^+ + {M_{N_{i}} \bar N_{L_i} N_{R_i}} + {\rm h.c.}, \label{eq:yukawa}
\\
{\cal V}&=
m^2_{\Phi} \Phi^\dag\Phi + {m^2_{\eta}} |\eta|^2 + m^2_\chi |\chi|^2 + m^2_{\chi^\pm} |\chi^+|^2 \nn\\
 & + \mu (\eta^T(i\sigma_2)\Phi\chi^- +{\rm h.c.}) +\mu_{\eta\chi}(\Phi^\dag\eta\chi^*+{\rm h.c.}) 
 + \mu_\chi (\chi^3+{\rm h.c.}) \nn\\
&  +\lambda_\Phi (\Phi^\dag\Phi)^2 
 +\lambda_0(\Phi^\dag\eta\chi^2+{\rm h.c.})
 +  \lambda_\eta (\eta^\dag\eta)^2  + \lambda_\chi (\chi^*\chi)^2 + \lambda_{\chi^\pm} (\chi^+\chi^-)^2 
+\lambda_{\Phi\eta} (\Phi^\dag\Phi)(\eta^\dag\eta)\nn\\
&+\lambda'_{\Phi\eta} |\Phi^\dag\eta|^2 
+\lambda_{\Phi\chi} (\Phi^\dag\Phi) \chi^*\chi +\lambda_{\eta\chi}(\eta^\dag\eta)  \chi^*\chi\nn\\
&+\lambda_{\Phi\chi^\pm} (\Phi^\dag\Phi) \chi^+\chi^- +\lambda_{\eta\chi^\pm}(\eta^\dag\eta)  \chi^+\chi^-
+\lambda_{\chi\chi^\pm}( \chi^*\chi)  \chi^+\chi^-
, 
\label{eq:pot}
\end{align}
where $\sigma_2$ is the second Pauli matrix, $i, j=1-3$, 
and the first term of $\mathcal{L}_{Y}$ can generate the SM
charged-lepton masses $m_\ell\equiv y_\ell v/\sqrt2$ ($\ell=1-3$) after the electroweak symmetry breaking.
Both  $N_i$ and $e_i$ can be considered as  the mass eigenstates without loss of generality.
For simplicity we assume all the parameters in (\ref{eq:yukawa}) are real and positive. 

The scalar fields can be parameterized as follows: 
\begin{align}
&\Phi =\left[
\begin{array}{c}
0\\
\frac{v+\phi}{\sqrt2}
\end{array}\right],\quad 
\eta =\left[
\begin{array}{c}
\eta^+\\
{\eta^0}
\end{array}\right], \quad \chi, \quad \chi^\pm,
\label{component}
\end{align}
where $\eta^0$ and $\chi$ are complex inert neutral bosons, $\eta^\pm$ and $\chi^\pm$ are the singly charged bosons, $v\simeq 246$ GeV is VEV of the Higgs doublet, and $\phi$ is the SM
Higgs boson with mass $m_\phi\approx125.5$ GeV.
{To ensure the stability of DM, the following condition should be at least satisfied: 
\begin{align}
\mid \mu + \mu_{\eta \chi} + \mu_\chi \mid < \sqrt{\Lambda}(m_\Phi^2 + m_\eta^2 + m_\chi^2 + m_{\chi^\pm}^2)^{\frac12}, \ 
\Lambda \equiv \sum_{\rm i= all\ quartic\ couplings} \lambda_i. 
\end{align}
}

Notice here that we have mixing between inert bosons through $\mu_{\eta\chi}$ and $\mu$, 
the resulting mass eigenvalues  and their rotation matrices are obtained by
\begin{align}
&O_{H}^TM_{H} O_{H} = \left[\begin{array}{cc} m_{H_1} & 0 \\ 0 & m_{H_2}  \end{array}\right], \quad
\left[\begin{array}{c} \chi \\ {\eta^0} \end{array}\right] = 
O_H \left[\begin{array}{c} H_1 \\ H_2 \end{array}\right]=
\left[\begin{array}{cc} \cos\alpha & \sin\alpha \\ -\sin\alpha & \cos\alpha \end{array}\right]
\left[\begin{array}{c} H_1 \\ H_2 \end{array}\right],\\
&V_C^TM_{H^\pm} V_C = \left[\begin{array}{cc} m_{H_1^\pm} & 0 \\ 0 & m_{H_2^\pm}  \end{array}\right], \quad
\left[\begin{array}{c} \chi^\pm \\ \eta^\pm \end{array}\right] = 
V_C \left[\begin{array}{c} H_1^\pm \\ H_2^\pm \end{array}\right]=
\left[\begin{array}{cc} \cos\beta & \sin\beta \\ -\sin\beta & \cos\beta \end{array}\right]
\left[\begin{array}{c} H_1^\pm \\ H_2^\pm \end{array}\right],
\end{align}
where $(H_{1(2)},H_{1(2)}^\pm)$ and ($\alpha$, $\beta$) can be written in terms of the parameters in the scalar potential, 
and we use the short hand notation  {$s_{\alpha(\beta)}$ and $c_{\alpha(\beta)}$ for $\sin\alpha(\beta)$ and $\cos\alpha(\beta)$} 
below~\footnote{See ref.~\cite{Aoki:2014cja} for scalar mass spectra in more details. 
}.

{ 
{\it DM candidate}:
The lightest neutral scalar $H_1$ is our DM candidate.
Here we consider constraints on $H_1$.
As for the direct detection experiment, the dominant elastic scattering cross section comes from Z-boson portal through mixing and found as
\begin{align}
\sigma_{SI}\simeq\left(\frac{m_{H_1} m_p}{m_{H_1} + m_p}\right)^2\frac{2G_F^2 s_\alpha^4}{\pi}\left( 1-4 s_w^2 \right)^2,
\end{align}
where $m_p$ is a proton mass, $s_w^2\approx0.23$ is the Weinberg angle and $G_F$ is the Fermi constant.
{Note here that $s_\alpha^4$ comes from the kinetic term of $\eta$, $D_\mu \eta^\dagger D_\mu \eta$, where the covariant derivative $D_\mu$ 
includes the SM gauge boson $Z$. Since $Z$-boson couples only to the isospin doublet scalar, in the effective $H_1-H_1-Z$ coupling only $\eta$ component
of $H_1$ couples to $Z$-boson. This leads to $s^2_\alpha$ suppression in the effective coupling and $s^4_\alpha$ suppression in the cross section for
the DM scattering off the nuclei. 
}
In the experiment of LUX~\cite{Akerib:2016vxi}, the typical upper bound on the cross section is
$\sigma_{SI}\lesssim10^{-45}$ cm$^2$ at $m_{H_1}={\cal O}(100)$ GeV.
Then the required condition on $\alpha$ is given by~\footnote{{If we consider the contribution of the Higgs portal~\cite{Cline:2013gha,
      Kanemura:2010sh} to the direct detection, the constraint is given by $2\lambda_{\Phi\chi} c_\alpha^2-2\frac{\mu_{\eta\chi}}{v}s_\alpha
    c_\alpha+(\lambda_{\Phi\eta}+\lambda'_{\Phi\eta})s^2_\alpha\lesssim 10^{-2}$. Thus one can satisfy the bound of direct detection by tuning
    the Higgs trilinear and quartic couplings.} }
\begin{align}
|s_\alpha| \lesssim 5 \times 10^{-2}.
\end{align}
It implies that the dominant component of the DM candidate is the gauge singlet boson $\chi$.
Hereafter we will neglect any terms proportional to $s_\alpha^4$. To explain the relic density, we rely on the resonant effect via s-channel of the
SM-Higgs, and we consider only the annihilation processes, since our DM is gauge singlet\footnote{{If our DM is dominated by $SU(2)_L$
    gauge doublet, then coannihilation is important and the allowed mass is at around 535 GeV~\cite{Hambye:2009pw}.}}.
In this case, to satisfy the relic density $\Omega h^2\approx0.12$, {the DM mass should be around the Higgs resonance region
 $m_{H_1} \approx m_\phi/2\approx63$ GeV~\cite{Kanemura:2010sh}. }

}

\subsection{Neutrino masses}
\begin{figure}[t] 
\begin{center}
\includegraphics[width=70mm]{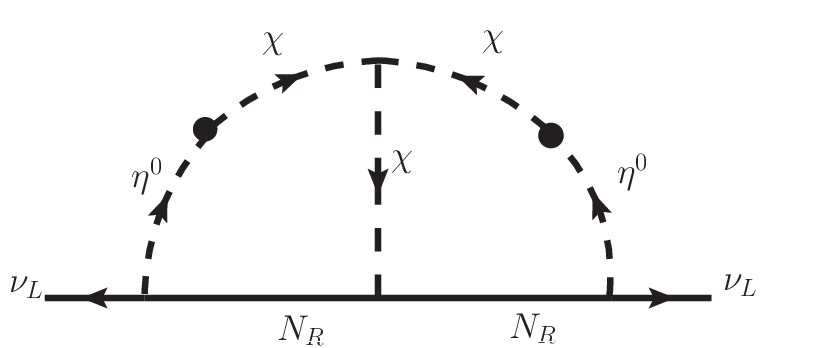}  
\caption{{Two-loop diagram to induce  neutrino mass matrix. Here} the blobs indicate the scalar mixing between $\eta^0$ and $\chi$.} 
\label{fig:neut1}
\end{center}
\end{figure}

The active neutrino mass matrix $m_\nu$  is generated at two-loop level as shown in {fig.}~\ref{fig:neut1}, 
and its formula is given by 
\begin{align}
&(m_{\nu})_{ij}
=
\frac{12\sqrt2 \mu_\chi s_\alpha^2 c_\alpha^2 (y_\eta)_{ia}(y_N)_{ab} (y_\eta)^T_{bj} }{(4\pi)^4} F_{II}
\equiv (y_\eta)_{ia}(R_N)_{ab} (y_\eta)^T_{bj},\\
&F_{II}=
\int \frac{[dx]}{z-1}\int[da]
 \left[c_\alpha^2\left[\ln\left(\frac{\Delta_{111}}{\Delta_{112}}\right) - \ln\left(\frac{\Delta_{211}}{\Delta_{212}}\right) \right] +
 s_\alpha^2\left[\ln\left(\frac{\Delta_{222}}{\Delta_{221}}\right) - \ln\left(\frac{\Delta_{122}}{\Delta_{121}}\right) \right] 
 \right],\\
& \Delta_{\ell mn}=-a\frac{x M_{N_b}^2 + y m_{H_n}^2 + z m_{H_m}^2}{z^2-z}+b M_{N_a}^2 + c m_{H_\ell}^2,
\end{align}
where {$R_N$ is a parameter with a mass dimension  and depends on the parameters $(y_N, \mu_\chi, s_\alpha, F_{II})$,}  
$ [dx] \equiv dx \, dy\,dz\,\delta(x+y+z-1)$, $ [da] \equiv da \, db \, dc\, \delta(a+b+c-1)$, and
we assume $y_N\equiv y_{N_R}\approx y_{N_L}$.
{As we mentioned in Section~\ref{sec:intro}, although they are generated at two-loop level,
the neutrino masses scale like three-loop model due to $s_\alpha$ suppression.}
{Then the active neutrino mass matrix $({m}_\nu)_{ij}$ can be} diagonalized by the Pontecorvo-Maki-Nakagawa-Sakata  (PMNS) mixing matrix $V_{\rm MNS}$~\cite{Maki:1962mu} as
{\begin{align}
({m}_\nu)_{ij} &=(V_{\rm MNS}^* D_\nu V_{\rm MNS}^\dag)_{ij},\quad D_\nu\equiv  \text{diag}(m_{\nu_1},m_{\nu_2},m_{\nu_3}),
\\
V_{\rm MNS}&=
\left[\begin{array}{ccc} {c_{13}}c_{12} &c_{13}s_{12} & s_{13} e^{-i\delta}\\
 -c_{23}s_{12}-s_{23}s_{13}c_{12}e^{i\delta} & c_{23}c_{12}-s_{23}s_{13}s_{12}e^{i\delta} & s_{23}c_{13}\\
  s_{23}s_{12}-c_{23}s_{13}c_{12}e^{i\delta} & -s_{23}c_{12}-c_{23}s_{13}s_{12}e^{i\delta} & c_{23}c_{13}\\
  \end{array}
\right].
\end{align}
We assume the neutrino masses are normal ordered, neglect the Majorana phases, and fix the Dirac phase $\delta=-\pi/2$ in the numerical analysis for simplicity.
Then we  apply the generalized Casas-Ibarra parametrization~\footnote{In our case, the central matrix is not diagonal but the symmetric matrix which
  is proportional to $y_N$.} to our analysis which use the observed neutrino oscillation data with global fit~\cite{Forero:2014bxa}. We impose
$\sum_{i=1-3}m_{\nu_i}< {0.12}$ eV at 95\% C.L. as reported by Planck collaboration~ {\cite{Aghanim:2018eyx}.}
The Yukawa coupling $y_\eta$ can be rewritten in terms of  the following parameters:
\begin{align}
y_\eta\approx V_{\rm MNS}^* \sqrt{D_\nu} {\cal O}(\theta_{i})\left({R_N}^{Ch}\right)^{-1}, 
\end{align}
where ${\cal O}(\theta_i)$ is an arbitrary orthogonal  $3\times3$ matrix with three complex values $\theta_{i}$ (i=1-3) that satisfy ${\cal O}{\cal O}^T={\cal O}^T{\cal O}={\rm Diag}(1,1,1)$
\footnote{In case  the family number of $N_R$ is two, ${\cal O}(\theta)$ is an arbitrary $3\times2$ matrix with a complex value $\theta$ that
  satisfies ${\cal O}{\cal O}^T={\rm Diag}(0,1,1)$ and ${\cal O}^T{\cal O}={\rm Diag}(1,1)$. The numerical result for the three families does not
  change much from this case.}
}
and $R_N^{ Ch}$ is Cholesky decomposed matrix. 
This matrix is a lower triangular matrix and satisfies the following relation (see appendix A), 
\begin{eqnarray}
R_N=R_N^{ Ch} \left( R_N^{ Ch} \right)^T. 
\end{eqnarray}


\subsection{Lepton flavor violations and muon $(g-2)$}

{\it  Lepton flavor violations(LFVs)}  at one-loop level arise from the terms with $y_\eta$ and $y_\chi$ 
{as shown in the left panel of fig. \ref{LFV}}~\footnote{Recently sophisticated analysis has been done by refs.~\cite{Lindner:2016bgg, Guo:2017ybk}}.
{Between two classes of LFV modes $\ell_i\to\ell_j\gamma$ and $\ell_i\to\ell_j\ell_k\ell_\ell$, the former tends to give more stringent
  bounds on the related couplings and masses~\cite{Toma:2013zsa}. Thus we consider only this mode below.}  The model prediction for the radiative
decay channel {in our case} is given by
\begin{align}
& {\rm BR}(\ell_i\to\ell_j\gamma)= \frac{48\pi^3C_{ij} \alpha_{\rm em} }{{\rm G_F^2}}
\left(|(a_{R}^{\eta\chi})_{ij} + a_{R_{ij}}^\eta + {\epsilon_{ij} a_{R_{ij}}^\chi}|^2
+| (a_{L}^{\eta\chi})_{ij}+ {\epsilon_{ij} a_{L_{ij}}^\eta + a_{L_{ij}}^\chi} |^2\right),\\
\label{eq:aR}
& (a_{R}^{\eta\chi})_{ij} = (a_{L}^{\eta\chi})^\dag_{ij} =
-\frac{s_\beta c_\beta}{(4\pi)^2}\sum_{k=1,2,3}\frac{M_{N_k}}{m_{\ell_i}} {(y_\eta)_{jk}(y_\chi)_{ki} } 
{\left(F_1(M_{N_k}, m_{H_1^\pm})-F_1(M_{N_k}, m_{H_2^\pm}) \right)}
,\\
\label{eq:aR_eta}
& a_{R_{ij}}^\eta = a_{L_{ij}}^\eta =\sum_{k=1,2,3}\frac{(y_\eta)_{jk}(y_\eta^\dag)_{ki} }{(4\pi)^2} 
\left[ s_\beta^2 F_{lfv}(M_{N_k},m_{H_1^\pm}) + c_\beta^2F_{lfv}(M_{N_k},m_{H_2^\pm})\right],\\
\label{eq:aR_chi}
& a_{R_{ij}}^\chi = a_{L_{ij}}^\chi =\sum_{k=1,2,3}\frac{(y_\chi^\dag)_{jk}(y_\chi)_{ki} }{(4\pi)^2} 
\left[ c_\beta^2 F_{lfv}(M_{N_k},m_{H_1^\pm}) + s_\beta^2F_{lfv}(M_{N_k},m_{H_2^\pm})\right],\\
&{F_1(m_1, m_2)=\frac{m_1^2+m_2^2}{2(m_1^2-m_2^2)^2} - \frac{m_1^2 m_2^2}{(m_1^2-m_2^2)^3}\log\frac{m_1^2}{m_2^2}, 
}
\\
& F_{lfv}(m_a,m_b)=\frac{2 m_a^6 +3 m_a^4 m_b^2 -6 m_a^2 m_b^4 +6 m_b^6+12 m_a^4 m_b^2 \ln(m_b/m_a)}{12(m_a^2-m_b^2)^4},
\end{align}
where $\epsilon_{ij} \equiv (m_{\ell_j}/m_{\ell_i}) (\ll 1)$, $\eta^\pm$ is the singly charged component of $\eta$, ${\rm G_F}\approx 1.17\times10^{-5}$[GeV]$^{-2}$ is the Fermi constant, $\alpha_{\rm em}\approx1/137$ is the fine structure constant, $C_{21}\approx1$, $C_{31}\approx 0.1784$, and $C_{32}\approx 0.1736$.
{Note that in the limit $m_{\ell_i} \to 0$  Eq.~(\ref{eq:aR}) is divergent, but this mass comes from the total decay rate and
  $m_{\ell_i} \to 0$ limit is unphysical.} 
Experimental upper bounds are ${\rm BR}(\mu\to e\gamma)\lesssim 4.2\times10^{-13}$, ${\rm BR}(\tau\to e\gamma)\lesssim 3.3\times10^{-8}$, and ${\rm BR}(\tau\to \mu\gamma)\lesssim 4.4\times10^{-8}$~\cite{pdg, TheMEG:2016wtm}.
\begin{figure}[t]
 \centering
  \begin{tabular}{c}
   \begin{minipage}{0.45\hsize}
    \centering
    \includegraphics[width=6.5cm]{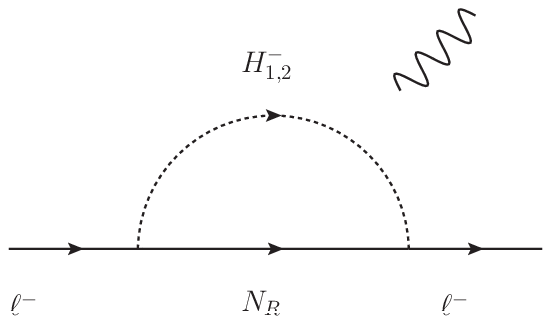}
    $\ell \to \ell \gamma$ and muon g-2
   \end{minipage}
   \begin{minipage}{0.45\hsize}
    \centering
    \includegraphics[width=5.5cm]{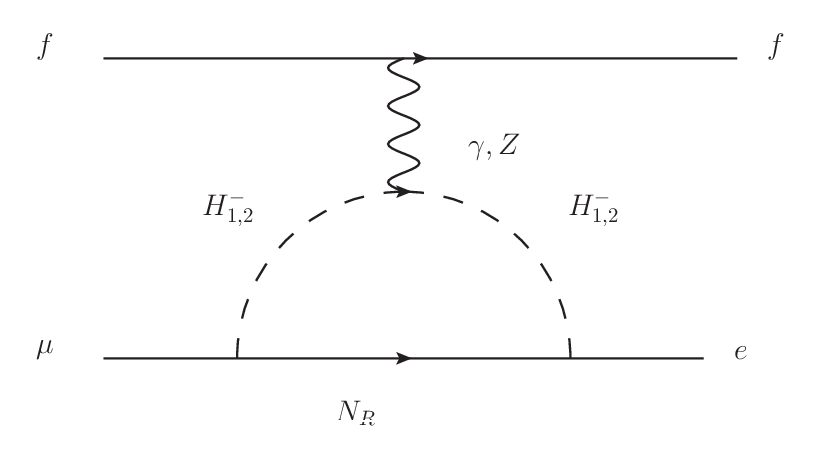}\\
    $\mu$-e conversion
   \end{minipage}
  \end{tabular}
 \caption{ 
 LFV diagrams
 }
 \label{LFV}
\end{figure}

The {\it $\mu-e$ conversion} rate can be expressed by using $a_{R/L}$ defined in Eqs.~(\ref{eq:aR}),(\ref{eq:aR_eta}), and (\ref{eq:aR_chi}). 
{The Feynman diagram is shown in the right panel of fig. \ref{LFV}}, and its capture rate $R$
is obtained approximately to be~\footnote{We neglect the contribution from the Z-penguin diagram due to suppression factor $(m_\ell/M_N)^2$.}
\begin{align}
&R\approx \frac{C_{\mu e}|Z|^2}{\Gamma_{\rm cap}}  
 {\left(|(a_{R}^{\eta\chi})_{\mu e} + a_{R_{\mu e}}^\eta +{\epsilon_{\mu e} a_{R_{\mu e}}^\chi}|^2
+| (a_{L}^{\eta\chi})_{\mu e}+ {\epsilon_{\mu e} a_{L_{\mu e}}^\eta + a_{L_{\mu e}}^\chi} |^2\right)},\
 C_{\mu e}\approx 4\alpha_{\rm em}^5 \frac{Z_{\rm eff}^4|F(q)|^2m_\mu^5}{Z},
\end{align}
where $Z$, $Z_{\rm eff}$, $F(q)$, and $R$ are  given in table~\ref{tab:mue-conv}.

Here let us define $Y\equiv \frac{R}{ {\rm BR}(\mu\to e\gamma)}$, since their flavor structures are same.
Then it is given by
\begin{align}
Y\approx 1.22\times 10^{-24}\left( \frac{Z Z_{\rm eff}^4 |F(q)|^2}{\Gamma_{\rm cap}}\right).
\end{align}
Depending on the nuclei,  $Y\approx {\cal O}$(0.1), as listed in  table~\ref{tab:mue-conv}.
It suggests that the constraint from the $\mu-e$ conversion is always satisfied once we satisfy the constraint of 
$\mu\to e \gamma$. 
We will discuss the {sensitivity of} future experiments, $R_{Ti}$ and $R_{Al}$, in the numerical analysis.

\begin{table}[t]
\begin{tabular}{c|c|c|c|c|c} \hline
Nucleus $^A_Z N$ & $Z_{\rm eff}$ & $|F(-m^2_\mu)|$ & $|\Gamma_{\rm capt}(10^6{\rm sec}^{-1})$ & Experimental bound (Future bound) &$Y\equiv \frac{R}{{\rm BR}(\mu\to e\gamma)}$ \\ \hline
$^{27}_{13} Al$ & $11.5$ & $0.64$ & $0.7054$  & ($R_{Al}\lesssim10^{-16}$)~\cite{Hungerford:2009zz} &$0.25$\\
$^{48}_{22} Ti$ & $17.6$ & $0.54$ &$2.59$  & 
$R_{Ti}\lesssim 4.3\times 10^{-12}$~\cite{Dohmen:1993mp}\ 
($\lesssim 10^{-18}$ \cite{Barlow:2011zza})  &$0.44$  \\
$^{197}_{79} Au$ & $33.5$ & $0.16$ & $13.07$ & $R_{Au}\lesssim7\times 10^{-13}$ ~\cite{Bertl:2006up} &$0.36$ \\ 
$^{208}_{82} Pb$ & $34$ & $0.15$ & $13.45$   & $R_{Pb}\lesssim4.6\times 10^{-11}$~\cite{Honecker:1996zf} &$0.34$ \\ \hline
\end{tabular}
\caption{
Summary for the the $\mu\mathchar`-e$ conversion in various nuclei: 
$Z$, $Z_{\rm eff}$, $F(q)$, $\Gamma_{\rm capt}$, and the bounds on
the capture rate $R$.}
\label{tab:mue-conv}
\end{table}
{\it New contribution to the muon anomalous magnetic moment (muon $g-2$)}{, whose diagram is displayed in the left panel of
  fig. \ref{LFV}}, is given by
\begin{align}
\Delta a_\mu\approx -m_\mu^2
{[(a_{R}^{\eta\chi})_{\mu \mu} + a_{R_{\mu \mu}}^\eta +{\epsilon_{\mu \mu} a_{R_{\mu \mu}}^\chi}
+ (a_{L}^{\eta\chi})_{\mu \mu}+ {\epsilon_{\mu \mu} a_{L_{\mu \mu}}^\eta + a_{L_{\mu \mu}}^\chi}]},
\end{align}
where only the terms $(a_{R(L)}^{\eta\chi})_{\mu\mu}$ are positive contributions to the muon $g-2$.
Thus we expect  $a_{R(L)_{\mu\mu}}^\eta, a_{R(L)_{\mu\mu}}^\chi \ll (a_{R/L}^{\eta\chi})_{\mu\mu}$ 
to explain the discrepancy between the experimental results and the theoretical predictions which is of order 
of ${\cal O}(10^{-9})$~\cite{Agashe:2014kda}. 

{{\it Lepton Flavor-Changing/Conserving  $Z$ Boson Decay}:
Here, we consider the flavor changing/conserving $Z$ boson decay $Z\to \ell_i^- \ell_j^+$ {as shown in fig. \ref{zdecay}}, whose 
branching fractions have been measured or restricted by experiments as~\cite{pdg}, 
\begin{align}
&{\rm BR}(Z\to e^- e^+)=(3.363\pm0.004)\ {\rm \%},\label{eq:z2mu-exp_conserve}\\
&{\rm BR}(Z\to \mu^- \mu^+)=(3.366\pm0.007)\  {\rm \%}, \\
&{\rm BR}(Z\to \tau^- \tau^+)=(3.370\pm0.008) \  {\rm \%}\\
&{\rm BR}(Z\to e^\mp \mu^\pm)\lesssim 7.5\times10^{-7}\ ,\\
&{\rm BR}(Z\to e^\mp \tau^\pm)\lesssim 9.8\times10^{-6}\,\\
&{\rm BR}(Z\to \tau^\mp \mu^\pm)\lesssim 1.2\times10^{-5}.\label{eq:z2mu-exp_change}
\end{align}
They will be improved by future experiments Giga-Z, ILC, and CEPC.
\begin{figure}[t]
 \centering
    \includegraphics[width=10cm]{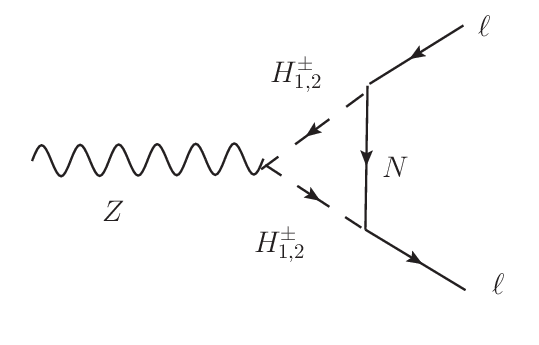}
 \caption{ 
The diagram for the lepton flavor-changing/conserving $Z$ boson decay}
 \label{zdecay}
\end{figure}
The model prediction for the branching fraction is
\begin{align}
\text{BR}(Z\to \ell_i^- \ell_j^+)
&=\frac{\Gamma(Z\to \ell_i^- \ell_j^+)}{\Gamma_{\rm tot}}
=\frac{m_Z}{24\pi \Gamma_{\rm tot}}(|\Gamma_{L_{ij}} |^2+|\Gamma_{R_{ij}}|^2)
~,\\
\Gamma_{L_{ij}} &\approx\frac{g_2}{c_w}\left(-\frac12+s_w^2\right)
\left[\delta_{ij}+\sum_{a=1-3}\frac{y_{\eta_{ia}} y^\dag_{\eta_{aj}}}{(4\pi)^2} G(M_{N_a},m_{H_2}^\pm)\right]~,\\
\Gamma_{R_{ij}} &\approx\frac{g_2 s_w^2}{c_w}
\left[\delta_{ij}+\sum_{a=1-3}\frac{y^\dag_{\chi_{ia}} y_{\chi_{aj}}}{(4\pi)^2} G(M_{N_a},m_{H_1}^\pm)\right]
\label{eq:Z2mu},
\end{align}
where we have neglected terms proportional to $s_\alpha^2$ and/or $(m_\ell/m_Z)^2$, 
$\Gamma_{\rm tot}\approx(2.4952\pm0.0023)$ GeV, 
and defined~\cite{Mohr:2015ccw}
\begin{align*}
G(m_a,m_b) &\approx\frac{m_a^4-4m_a^2m_b^2+3m_b^4-4 m_a^4\ln[m_a]+8m_a^2m_b^2\ln[m_a]-4m_b^4\ln[m_b]}
{4(m_a^2-m_b^2)^2} ~.
\end{align*} 
They are constrained by Eqs.~(\ref{eq:z2mu-exp_conserve}) $-$(\ref{eq:z2mu-exp_change}).
}

\section{Numerical analysis \label{sec:numerical}}
{For the numerical analysis, we generate input parameters randomly  in the following ranges:
\begin{align}
& 
(y_{N_{ij}},|y_\chi|)  \in  [10^{-8}, 0.1],\ \theta_{1,2,3} \in [10^{-3}i,2\pi+100i],\  s_\alpha \in [10^{-5}, 10^{-3}],\ 
s_\beta \in [-1, 1],
\\
& (\mu, \mu_{\chi},\mu_{\eta\chi}) \in [10^3]\ {\rm GeV}, \  m_{H_{1,2}^\pm} \in [{80},10^3]\ {\rm GeV}, \ 
(m_{H_2}, M_{N_1},M_{N_2},M_{N_3} ) \in [200, 10^3] \ {\rm GeV}, \nn
\end{align}
}
where $i,j=1,2,3$, $y_N$ is a symmetric matrix, $\theta_{1,2,3}$ are arbitrary complex values in the Casas-Ibarra {parametrization}.
We fixed $m_{H_1}=m_\phi/2$.
{The lower bound on $H^{\pm}_{1,2}$, 80 GeV, comes from the LEP experiment~\cite{pdg, Barate:2003sz}.
In addition, the LHC gives a mass bound for the charged boson.
Especially,  the $SU(2)_L$ originated charged boson would have a  feature similar to the slepton in the supersymmetric model, since it decays into
a charged lepton and missing energy. 
{The lower bound of the mass from the CMS collaboration is 450 GeV~\cite{Sirunyan:2018nwe}.}
Hence we might apply this bound for our case, although the detail analysis is beyond our scope of this paper.
}
Notice here  $y_\eta$ should satisfy the  perturbative limit; $y_\eta\lesssim \sqrt{4\pi}$.
}

{In fig.~\ref{fig:mu_e_gamma}, we show scatter plots  BR($\tau\to e\gamma$)(red) and BR($\tau\to\mu\gamma$)(blue) 
as a function of BR($\mu \to e \gamma$).
It suggests that BR($\tau\to e\gamma$) and BR($\tau\to\mu\gamma$) are much less than the upper bounds of experiments, while the maximum value of BR($\mu\to e\gamma$) reaches the experimental upper bound. 
In fig.~\ref{fig:mu_g2}, we show scatter plots of BR$(Z\to e\tau)$(red) and BR$(Z\to\mu\tau)$(blue) as a function of BR$(Z \to e\mu)$.
It suggests that BR($Z\to e\tau$) and BR($Z\to\mu\tau$) are much less than the upper bounds of experiments, while the maximum value of 
BR$(Z \to e\mu)$ is close to the experimental upper bound. Thus BR$(Z \to e\mu)$ could be tested in the future experiments.
As for muon $g-2$, the maximum value is at most 5$\times10^{-15}$, which is much smaller than the current discrepancy.
This is because the positive contribution comes from the mixing term between $y_\eta$ and $y_\chi$ only. 


\begin{figure}[t]
\begin{center}
\includegraphics[width=10cm]{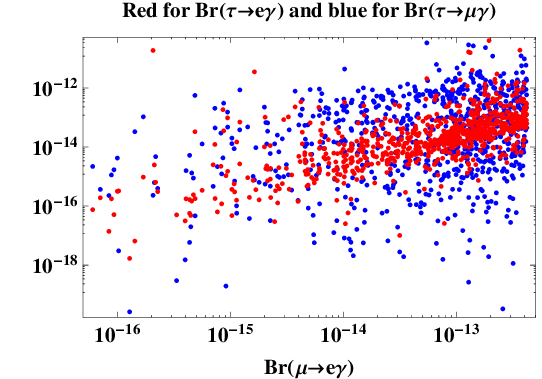}
\caption{ 
 Scatter plots of BR($\tau\to e\gamma$)(red) and BR($\tau\to\mu\gamma$)(blue) in terms of BR($\mu \to e \gamma$).
} \label{fig:mu_e_gamma}
\end{center}\end{figure}

\begin{figure}[t]
\begin{center}
\includegraphics[width=10cm]{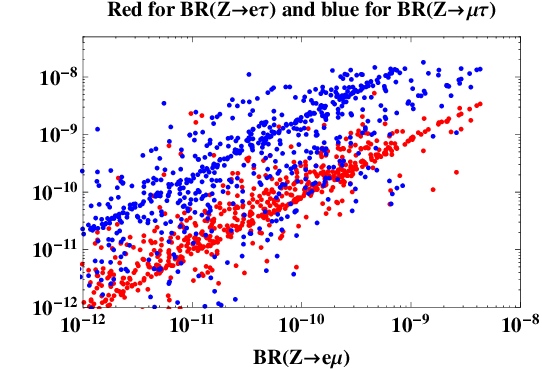}
\caption{ 
Scatter plots of BR$(Z\to e\tau)$(red) and BR$(Z\to\mu\tau)$(blue) in terms of BR$(Z \to e\mu)$.} \label{fig:mu_g2}
\end{center}\end{figure}

\section{Summaries and discussions}
We have explored the possibility to explain bosonic dark matter candidate with {a} gauge singlet inside the loop to generate the neutrino mass matrix at two-loop level.
{Here, our setup is} the Zee-Babu type scenario with $Z_3$ discrete symmetry, in which we have considered the neutrino oscillation data, DM, and lepton flavor violations. 

First of all, we have found the upper bound on $s_\alpha$ to be of the order $10^{-2}$ from the direct detection experiment.
Thus the only solution to satisfy the observed relic density is to use the SM Higgs resonance with the DM mass 
around  the half of the Higgs mass, $m_{H_1}\approx m_\phi/2$.

Second, the neutrino mass matrix is reduced by not only the two-loop suppression but also $s_\alpha^2\approx10^{-3}$ suppression that comes from
{the direct detection bound of} the DM. As a result, the scale of the neutrino masses is equivalent to that of the three-loop neutrino model.

{We have found that the positive muon $g-2$ thanks to $\chi^\pm$.
But its typical value  ${\cal O}(10^{-14})$ in our global analysis is too small to explain the $\sim 3\sigma$
discrepancy of the muon $g-2$ between the experiment and the SM.}


{We briefly mention the possibility to detect our new particles at the LHC or the ILC.
In these kinds of radiative seesaw models, they tend to have large Yukawa couplings in the lepton sector.
Therefore, the effect of LFV processes can be large.
On the other hand, 
the bounds from the LHC experiments are typically weaker than the LFV constraints, since
new scalar bosons do not couple to quarks.
Hence, in this paper we consider only LFV effects.}


}


\section*{Acknowledgments}
\vspace{0.5cm}
This work was supported in part by the National Research Foundation of Korea(NRF) grant funded by
the Korea government(MSIT), Grant No. NRF-2018R1A2A3075605 (S.B.). 
H. O. is sincerely grateful for all the KIAS members in  Korea.
This research is supported by the Ministry of Science, ICT and Future Planning, Gyeongsangbuk-do and Pohang City (H.O.).
The work was supported from 
European Regional Development Fund-Project 
Engineering Applications of Microworld Physics 
(No. CZ.02.1.01/0.0/0.0/16-019/0000766) (Y.O.). 


\begin{appendix}
\section{Cholesky decomposition}
A symmetric matrix M can be factorized by Cholesky decomposition.
The decomposition is as follows:  
\begin{eqnarray}
M=
\left[ 
\begin{array}{ccc}
m_{11} &m_{12} &m_{13} \\
m_{12}&m_{22} &m_{23} \\
m_{13}&m_{23} & m_{33} \\
\end{array} 
\right]
=L L^T, 
\end{eqnarray}
where L is a lower triangular matrix. 
The explicit form for the matrix L is 
\begin{eqnarray}
L=
\left[ 
\begin{array}{ccc}
l_{11}&0 &0 \\
l_{21}&l_{22} &0 \\
l_{31}&l_{32} &l_{33} \\
\end{array} 
\right], 
\end{eqnarray}
\begin{eqnarray}
l_{11}=\pm\sqrt{m_{11}}, \\
l_{21}=\frac{m_{12}}{l_{11}}, \\
l_{22}=\pm\sqrt{m_{22}-l_{21}^2}, \\
l_{31}=\frac{m_{13}}{l_{11}}, \\
l_{32}=\frac{m_{32}-l_{31}l_{21}}{l_{22}}, \\
l_{33}=\pm\sqrt{m_{33}-l_{31}^2-l_{32}^2}, 
\end{eqnarray}
where we assumed all parameters to be real and positive.

\end{appendix}


\begin{thebibliography}{99}

\bibitem{2-lp-zB} 
 A.~Zee,
 Nucl.\ Phys.\ B {\bf 264}, 99  (1986);
 K.~S.~Babu,
 Phys.\ Lett.\ B {\bf 203}, 132  (1988).
 
 
\bibitem{Babu:2002uu} 
  K.~S.~Babu and C.~Macesanu,
  Phys.\ Rev.\ D {\bf 67}, 073010 (2003)
  [hep-ph/0212058].

\bibitem{AristizabalSierra:2006gb} 
  D.~Aristizabal Sierra and M.~Hirsch,
  JHEP {\bf 0612}, 052 (2006)
  [hep-ph/0609307].

\bibitem{Nebot:2007bc} 
  M.~Nebot, J.~F.~Oliver, D.~Palao and A.~Santamaria,
  Phys.\ Rev.\ D {\bf 77}, 093013 (2008)
  [arXiv:0711.0483 [hep-ph]].

\bibitem{Schmidt:2014zoa} 
  D.~Schmidt, T.~Schwetz and H.~Zhang,
  Nucl.\ Phys.\ B {\bf 885}, 524 (2014)
  [arXiv:1402.2251 [hep-ph]].

\bibitem{Herrero-Garcia:2014hfa} 
  J.~Herrero-Garcia, M.~Nebot, N.~Rius and A.~Santamaria,
  Nucl.\ Phys.\ B {\bf 885}, 542 (2014)
  [arXiv:1402.4491 [hep-ph]].

\bibitem{Long:2014fja} 
  H.~N.~Long and V.~V.~Vien,
  Int.\ J.\ Mod.\ Phys.\ A {\bf 29}, no. 13, 1450072 (2014)
  [arXiv:1405.1622 [hep-ph]].

\bibitem{VanVien:2014apa} 
  V.~Van Vien, H.~N.~Long and P.~N.~Thu,
  arXiv:1407.8286 [hep-ph].

\bibitem{Aoki:2010ib} 
  M.~Aoki, S.~Kanemura, T.~Shindou and K.~Yagyu,
  JHEP {\bf 1007}, 084 (2010)
  [Erratum-ibid.\  {\bf 1011}, 049 (2010)]
  [arXiv:1005.5159 [hep-ph]].

\bibitem{Lindner:2011it} 
  M.~Lindner, D.~Schmidt and T.~Schwetz,
  Phys.\ Lett.\ B {\bf 705}, 324 (2011)
  [arXiv:1105.4626 [hep-ph]].

\bibitem{Baek:2012ub} 
  S.~Baek, P.~Ko, H.~Okada and E.~Senaha,
  JHEP {\bf 1409}, 153 (2014)
  [arXiv:1209.1685 [hep-ph]].

\bibitem{Aoki:2013gzs} 
  M.~Aoki, J.~Kubo and H.~Takano,
  Phys.\ Rev.\ D {\bf 87}, no. 11, 116001 (2013)
  [arXiv:1302.3936 [hep-ph]].

\bibitem{Kajiyama:2013zla} 
  Y.~Kajiyama, H.~Okada and K.~Yagyu,
  Nucl.\ Phys.\ B {\bf 874}, 198 (2013)
  [arXiv:1303.3463 [hep-ph]].

\bibitem{Kajiyama:2013rla} 
  Y.~Kajiyama, H.~Okada and T.~Toma,
  Phys.\ Rev.\ D {\bf 88}, 015029 (2013)
  [arXiv:1303.7356].

\bibitem{Baek:2013fsa} 
  S.~Baek, H.~Okada and T.~Toma,
  JCAP {\bf 1406}, 027 (2014)
  [arXiv:1312.3761 [hep-ph]].

\bibitem{Okada:2014vla} 
  H.~Okada,
  arXiv:1404.0280 [hep-ph].

\bibitem{Okada:2014qsa} 
  H.~Okada, T.~Toma and K.~Yagyu,
  Phys.\ Rev.\ D {\bf 90}, no. 9, 095005 (2014)
  [arXiv:1408.0961 [hep-ph]].

\bibitem{Okada:2015nga} 
  H.~Okada,
  arXiv:1503.04557 [hep-ph].

\bibitem{Geng:2015sza} 
  C.~Q.~Geng and L.~H.~Tsai,
  arXiv:1503.06987 [hep-ph].

\bibitem{Kashiwase:2015pra} 
  S.~Kashiwase, H.~Okada, Y.~Orikasa and T.~Toma,
  Int.\ J.\ Mod.\ Phys.\ A {\bf 31}, no. 20n21, 1650121 (2016)
  doi:10.1142/S0217751X16501219
  [arXiv:1505.04665 [hep-ph]].
  
  
\bibitem{Aoki:2014cja} 
  M.~Aoki and T.~Toma,
  JCAP {\bf 1409}, 016 (2014)
  [arXiv:1405.5870 [hep-ph]].
  


\bibitem{Baek:2014awa} 
  S.~Baek, H.~Okada and T.~Toma,
  Phys.\ Lett.\ B {\bf 732}, 85 (2014)
  [arXiv:1401.6921 [hep-ph]].

\bibitem{Okada:2015nca} 
  H.~Okada and Y.~Orikasa,
  Phys.\ Rev.\ D {\bf 93}, no. 1, 013008 (2016)
  doi:10.1103/PhysRevD.93.013008
  [arXiv:1509.04068 [hep-ph]].
  
  
  
\bibitem{Sierra:2014rxa} 
  D.~Aristizabal Sierra, A.~Degee, L.~Dorame and M.~Hirsch,
  JHEP {\bf 1503}, 040 (2015)
  [arXiv:1411.7038 [hep-ph]].

\bibitem{Nomura:2016rjf} 
  T.~Nomura and H.~Okada,
  Phys.\ Lett.\ B {\bf 756}, 295 (2016)
  [arXiv:1601.07339 [hep-ph]].

\bibitem{Nomura:2016run} 
  T.~Nomura, H.~Okada and Y.~Orikasa,
  arXiv:1602.08302 [hep-ph].

\bibitem{Bonilla:2016diq} 
  C.~Bonilla, E.~Ma, E.~Peinado and J.~W.~F.~Valle,
  arXiv:1607.03931 [hep-ph].

\bibitem{Kohda:2012sr} 
  M.~Kohda, H.~Sugiyama and K.~Tsumura,
  Phys.\ Lett.\ B {\bf 718}, 1436 (2013)
  [arXiv:1210.5622 [hep-ph]].

  \bibitem{Dasgupta:2013cwa} 
  B.~Dasgupta, E.~Ma and K.~Tsumura,
  Phys.\ Rev.\ D {\bf 89}, 041702 (2014)
  [arXiv:1308.4138 [hep-ph]].

\bibitem{Baek:2014sda} 
  S.~Baek,
  JHEP {\bf 1508}, 023 (2015)
  doi:10.1007/JHEP08(2015)023
  [arXiv:1410.1992 [hep-ph]].

\bibitem{Nomura:2016ask} 
  T.~Nomura and H.~Okada,
  Phys.\ Rev.\ D {\bf 94}, 075021 (2016)
  doi:10.1103/PhysRevD.94.075021
  [arXiv:1607.04952 [hep-ph]].
  
  
\bibitem{Nomura:2016pgg} 
  T.~Nomura and H.~Okada,
  arXiv:1609.01504 [hep-ph].
  
  
\bibitem{Nomura:2016dnf} 
  T.~Nomura, H.~Okada and Y.~Orikasa,
  Phys.\ Rev.\ D {\bf 94}, no. 11, 115018 (2016)
  doi:10.1103/PhysRevD.94.115018
  [arXiv:1610.04729 [hep-ph]].
  
  
\bibitem{Liu:2016mpf} 
  Z.~Liu and P.~H.~Gu,
  arXiv:1611.02094 [hep-ph].
  


  
\bibitem{Liang:2016pvm} 
  Y.~F.~Liang {\it et al.},
  Phys.\ Rev.\ D {\bf 93}, no. 10, 103525 (2016)
  doi:10.1103/PhysRevD.93.103525
  [arXiv:1602.06527 [astro-ph.HE]].
  
\bibitem{Barlow:2011zza} 
  R.~J.~Barlow,
  Nucl.\ Phys.\ Proc.\ Suppl.\  {\bf 218}, 44 (2011).
  doi:10.1016/j.nuclphysbps.2011.06.009
  
  

\bibitem{Akerib:2016vxi} 
  D.~S.~Akerib {\it et al.},
  arXiv:1608.07648 [astro-ph.CO].
  
\bibitem{Cline:2013gha} 
  J.~M.~Cline, K.~Kainulainen, P.~Scott and C.~Weniger,
  Phys.\ Rev.\ D {\bf 88}, 055025 (2013)
  Erratum: [Phys.\ Rev.\ D {\bf 92}, no. 3, 039906 (2015)]
  doi:10.1103/PhysRevD.92.039906, 10.1103/PhysRevD.88.055025
  [arXiv:1306.4710 [hep-ph]].
    
\bibitem{Hambye:2009pw} 
  T.~Hambye, F.-S.~Ling, L.~Lopez Honorez and J.~Rocher,
  JHEP {\bf 0907}, 090 (2009)
  Erratum: [JHEP {\bf 1005}, 066 (2010)]
  doi:10.1007/JHEP05(2010)066, 10.1088/1126-6708/2009/07/090
  [arXiv:0903.4010 [hep-ph]].
  
\bibitem{Kanemura:2010sh} 
  S.~Kanemura, S.~Matsumoto, T.~Nabeshima and N.~Okada,
  Phys.\ Rev.\ D {\bf 82}, 055026 (2010)
  doi:10.1103/PhysRevD.82.055026
  [arXiv:1005.5651 [hep-ph]].
  
  
  
 

\bibitem{Maki:1962mu} 
  Z.~Maki, M.~Nakagawa and S.~Sakata,
  Prog.\ Theor.\ Phys.\  {\bf 28}, 870 (1962).
  doi:10.1143/PTP.28.870
  
  
\bibitem{Forero:2014bxa} 
  D.~V.~Forero, M.~Tortola and J.~W.~F.~Valle,
  Phys.\ Rev.\ D {\bf 90}, no. 9, 093006 (2014)
  doi:10.1103/PhysRevD.90.093006
  [arXiv:1405.7540 [hep-ph]].
  
  
\bibitem{Aghanim:2018eyx} 
  N.~Aghanim {\it et al.} [Planck Collaboration],
  arXiv:1807.06209 [astro-ph.CO].
 

\bibitem{Toma:2013zsa} 
  T.~Toma and A.~Vicente,
  JHEP {\bf 1401}, 160 (2014)
  doi:10.1007/JHEP01(2014)160
  [arXiv:1312.2840 [hep-ph]].

\bibitem{Ade:2015xua} 
  P.~A.~R.~Ade {\it et al.} [Planck Collaboration],
  Astron.\ Astrophys.\  {\bf 594}, A13 (2016)
  doi:10.1051/0004-6361/201525830
  [arXiv:1502.01589 [astro-ph.CO]].



\bibitem{Lindner:2016bgg} 
  M.~Lindner, M.~Platscher and F.~S.~Queiroz,
  arXiv:1610.06587 [hep-ph].

\bibitem{Guo:2017ybk} 
  C.~Guo, S.~Y.~Guo, Z.~L.~Han, B.~Li and Y.~Liao,
  arXiv:1701.02463 [hep-ph].
  
  
  
\bibitem{pdg} 
  C.~Patrignani {\it et al.} [Particle Data Group],
  Chin.\ Phys.\ C {\bf 40}, no. 10, 100001 (2016),

  M.~Tanabashi {\it et al.} [Particle Data Group],
  Phys.\ Rev.\ D {\bf 98}, no. 3, 030001 (2018).
  doi:10.1103/PhysRevD.98.030001
  
\bibitem{TheMEG:2016wtm} 
  A.~M.~Baldini {\it et al.} [MEG Collaboration],
  Eur.\ Phys.\ J.\ C {\bf 76}, no. 8, 434 (2016)
  [arXiv:1605.05081 [hep-ex]].


\bibitem{Agashe:2014kda} 
  K.~A.~Olive {\it et al.} [Particle Data Group],
  Chin.\ Phys.\ C {\bf 38}, 090001 (2014).
    
\bibitem{Mohr:2015ccw} 
  P.~J.~Mohr, D.~B.~Newell and B.~N.~Taylor,
  Rev.\ Mod.\ Phys.\  {\bf 88}, no. 3, 035009 (2016)
  doi:10.1103/RevModPhys.88.035009
  [arXiv:1507.07956 [physics.atom-ph]].

  
\bibitem{Hungerford:2009zz} 
  E.~V.~Hungerford [COMET Collaboration],
  AIP Conf.\ Proc.\  {\bf 1182}, 694 (2009).


\bibitem{Dohmen:1993mp} 
  C.~Dohmen {\it et al.} [SINDRUM II Collaboration],
  Phys.\ Lett.\ B {\bf 317}, 631 (1993).
  
  

\bibitem{Bertl:2006up} 
  W.~H.~Bertl {\it et al.} [SINDRUM II Collaboration],
  Eur.\ Phys.\ J.\ C {\bf 47}, 337 (2006).
  
\bibitem{Honecker:1996zf} 
  W.~Honecker {\it et al.} [SINDRUM II Collaboration],
  Phys.\ Rev.\ Lett.\  {\bf 76}, 200 (1996).

  

  
  
\bibitem{Griest:1990kh} 
  K.~Griest and D.~Seckel,
  Phys.\ Rev.\ D {\bf 43}, 3191 (1991).
  doi:10.1103/PhysRevD.43.3191
  
  
  
  
\bibitem{Ellis:1975ap} 
  J.~R.~Ellis, M.~K.~Gaillard and D.~V.~Nanopoulos,
  Nucl.\ Phys.\ B {\bf 106}, 292 (1976).
  doi:10.1016/0550-3213(76)90382-5
  
\bibitem{Shifman:1979eb} 
  M.~A.~Shifman, A.~I.~Vainshtein, M.~B.~Voloshin and V.~I.~Zakharov,
  Sov.\ J.\ Nucl.\ Phys.\  {\bf 30}, 711 (1979)
  [Yad.\ Fiz.\  {\bf 30}, 1368 (1979)].
  
\bibitem{Djouadi:2005gi} 
  A.~Djouadi,
  Phys.\ Rept.\  {\bf 457}, 1 (2008)
  doi:10.1016/j.physrep.2007.10.004
  [hep-ph/0503172].
  
\bibitem{Gastmans:2011wh} 
  R.~Gastmans, S.~L.~Wu and T.~T.~Wu,
  arXiv:1108.5872 [hep-ph].
  
\bibitem{Carena:2012xa} 
  M.~Carena, I.~Low and C.~E.~M.~Wagner,
  JHEP {\bf 1208}, 060 (2012)
  doi:10.1007/JHEP08(2012)060
  [arXiv:1206.1082 [hep-ph]].
  
\bibitem{Nishiwaki:2015iqa} 
  K.~Nishiwaki, H.~Okada and Y.~Orikasa,
  Phys.\ Rev.\ D {\bf 92}, no. 9, 093013 (2015)
  doi:10.1103/PhysRevD.92.093013
  [arXiv:1507.02412 [hep-ph]].
 
\bibitem{Hambye:2006zn} 
  T.~Hambye, K.~Kannike, E.~Ma and M.~Raidal,
  Phys.\ Rev.\ D {\bf 75}, 095003 (2007)
  doi:10.1103/PhysRevD.75.095003
  [hep-ph/0609228].

  
  
\bibitem{Barate:2003sz} 
  R.~Barate {\it et al.} [ALEPH and DELPHI and L3 and OPAL Collaborations and LEP Working Group for Higgs boson searches],
  Phys.\ Lett.\ B {\bf 565}, 61 (2003)
  doi:10.1016/S0370-2693(03)00614-2
  [hep-ex/0306033].
  
  
  \bibitem{Sirunyan:2018nwe} 
  A.~M.~Sirunyan {\it et al.} [CMS Collaboration],
  [arXiv:1806.05264 [hep-ex]].
  
\end{thebibliography}
\end{document}